\documentclass{iaus}
\usepackage{graphicx}
\usepackage{natbib}

\title[Jet launching and field advection] %% give here short title %%
{Jet launching and field advection in quasi-Keplerian discs}

\author[J. Ferreira \& P.-O. Petrucci]   %% give here short author list %%
{Jonathan Ferreira
\and Pierre Olivier Petrucci}

\affiliation{Laboratoire d'Astrophysique de Grenoble, CNRS,
             Universit\'e Joseph Fourier, B.P. 53, F-38041 Grenoble, France}
\pubyear{2010}
\volume{275}  %% insert here IAU Symposium No.
\jname{Jets at all scales}
\editors{Gustavo E. Romero, Rashid A. Sunyaev \& Tomaso Belloni, eds}
\begin{document}

\maketitle

\begin{abstract}

The fact that self-confined jets are observed around black holes, neutron stars and young forming stars points to a jet launching mechanism independent of the nature of the central object, namely the surrounding accretion disc. The properties of Jet Emitting Discs (JEDs) are briefly reviewed. It is argued that, within an alpha prescription for the turbulence (anomalous viscosity and diffusivity), the steady-state problem has been solved. Conditions for launching jets are very stringent and require  a large scale magnetic field $B_z$ close to equipartition with the total (gas and radiation) pressure. The total power feeding the jets decreases with the disc thickness: fat ADAF-like structures with $h\sim r$ cannot drive super-Alfv\'enic jets. However, there exist also hot, optically thin JED solutions that would be observationally very similar to ADAFs.

Finally, it is argued that variations in the large scale magnetic $B_z$ field is the second parameter required to explain hysteresis cycles seen in LMXBs (the first one would be $\dot M_a$). 

\end{abstract}

\section{Jet Emitting Discs}

Self-confined jets are observed around a wide variety of astrophysical objects, namely from young forming stars (YSOs), neutron stars or stellar black holes (X-ray binaries) and supermassive black holes (AGN). Although the underlying emission mechanisms are quite different (molecular and atomic lines in YSOs, synchrotron radiation from supra-thermal particles in LMXBs and AGN), they share common properties: a high degree of collimation (opening angle of a few degrees), a systematic link with the accretion disc and carry away a sizable fraction of the released accretion power. It is thus natural to seek for a model where accretion and ejection processes are interdependent {\em regardless of the central object} (be it a star or a compact object).   

The most promising and so far developed model is ejection from a quasi-Keplerian accretion disc, commonly referred to  \cite{blan82} jets (hereafter BP). In this model, a large scale magnetic field of bipolar topology is assumed to thread a near-Keplerian accretion disc. This field extracts angular momentum and energy from the disc and transfers them back to the ejected material. This material is then accelerated to infinity and maintained self-confined by the hoop-stress due to the strong toroidal field. This self-confinement effect has been shown by \cite{blan82} and discussed by many other subsequent authors. However, what remained to be done was to compute the connection with the underlying disc, since the BP model was using ad-hoc boundary conditions at the disc surface.

This work has been done more than 10 years ago by \cite{ferr93a,ferr95,ferr97}. Using a self-similar ansatz, it is indeed possible to solve exactly the {\em full set} of non-relativistic MHD equations describing a resistive, viscous accretion disc thread by a large scale field. Both viscosity $\nu_v$ and magnetic diffusivity $\nu_m$ are assumed to arise from a MHD turbulence and were modeled with an alpha-prescription, namely $\nu_m= \alpha_m V_A h$ where $V_A$ is the Alfv\'en velocity on the disc midplane, $h$ the disc vertical scale height and $\nu_v= {\cal P}_m \nu_m$ where ${\cal P}_m$ is the effective magnetic Prandtl number. The set of MHD equations have been solved from the disc midplane to the jet termination point, subjected to the constraint of smoothly crossing the critical points of the flow (mainly slow-magnetosonic and Alfv\'en). These two regularity conditions are unbiased by the self-similar ansatz and severely constrain the parameter space of these Jet Emitting Discs (hereafter JEDs). Indeed, defining the disc magnetization $\mu= B_z^2/P$ where $P=P_{gas}+P_{rad}$, it has been shown that powerful jets can only be obtained with $0.1 < \mu < 1$, namely a large scale field close to equipartition. This result was in severe contrast with previous works (e.g. \cite{ogil98,ogil01}) and explains why the latter could {\em not} obtain super-Alfvenic, steady-state jets from thin discs. Solutions linked to the disc and crossing the three MHD critical points were also obtained, but it was shown that self-similarity introduces a strong bias on the fast-magnetosonic point that is probably unphysical \citep{ferr04}. 

Because of the mass loss, the disc accretion rate varies as $\dot M_a \propto r^\xi$, where $\xi$ is the disc ejection efficiency: the higher $\xi$ the more mass is being loaded onto the jets.  Models with isothermal or adiabatic magnetic surfaces achieve typical values of $\xi \sim 0.01$ \citep{cass00a}, whereas more heavily loaded jets with $\xi \sim 0.1$ can be obtained when a surface heating is present \citep{cass00b}. The asymptotic velocity reached along a given field line is a direct function of $\xi$. In a non-relativistic jet, it writes $v_j = v_K \sqrt{2 \lambda -3}$, where $\lambda \simeq 1 + \frac{1}{2\xi}$ is the magnetic lever arm and $v_K$ the Keplerian speed at the line footpoint. Thus, contrary to the usual belief, jets from near-Keplerian accretion discs can have velocities much larger than $v_K$. However, because the mass loss is not that tiny, highly relativistic jets are also difficult. Indeed, the jet magnetization, namely the ratio of the MHD Poynting flux to the kinetic energy flux at the disc surface is $\sigma \simeq \xi^{-1}$ and never reaches the high values ($10^4$ to $10^6$) sometimes invoked. As a consequence, only mildly relativistic flows are expected, but not flows with bulk Lorentz factors around 10 or 20. For those, another mechanism must be found, either extracting energy from a rotating black hole \citep{blan77} or a two-component jet (see e.g. \citet{bout08} and references therein).   
  
\begin{figure}
\begin{center}
\includegraphics[width=0.7\textwidth]{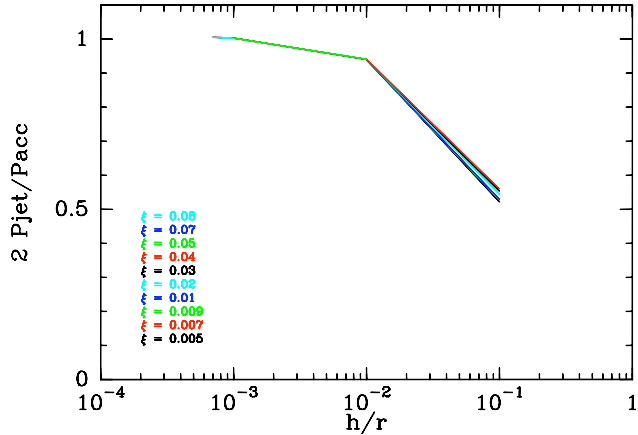}
\end{center}
\caption[]{Ratio of the total jet power to the total accretion power as a function of the disc thickness, for all isothermal models of \cite{ferr97}. Note that the power decreases rapidly, regardless of the value of the disc ejection efficiency $\xi$.}
\label{fig:Pjets}
\end{figure}

Although exact and solving the long standing problem of the jet launching from accretion discs, these solutions did not catch much attention in the AGN and LMXB communities. This is probably due to the fact that the ADAF solution (e.g. \citealt{nara94}) drew all the attention because of its success in explaining low luminosity hard spectra \citep{nara95}, while hoping to explain also jet formation \citep{nara95a}. Although no clear proof of this latter statement was ever produced, this idea remained deeply embedded in the two aforementioned communities. Self-similar JED models are actually able to address this question, since all terms are included in the calculations. Indeed, \cite{cass00a} remarked that they could not find any steady-state JED solution with a disc aspect ratio $\varepsilon= h/r > 0.3$. This is because, as the disc gets thicker, the effect of the radial magnetic tension increases drastically so as to cancel rotation at the disc surface. As a consequence, the toroidal field vanishes and so does the MHD Poynting flux: there is no more power to feed the jets. Figure~(\ref{fig:Pjets}) shows the total power $P_{jets}$ (mechanical, thermal and electromagnetic) leaving the disc normalized to the accretion power $P_{acc}$ as a function of the disc aspect ratio $h/r$. Geometrically thin discs give rise to jets that carry away up to 99\% of the released accretion power whereas slim discs with $\varepsilon =0.1$ can only provide 50\% to their jets. This power goes rapidly to zero so that no super-Alfv\'enic isothermal jet can be found from thick discs with $\varepsilon= h/r > 0.3$.  Hence, ADAFs are unable to drive powerful jets, although thermally driven winds would be a natural outcome (ADIOS, \citealt{blan99}). Global MHD simulations do show ejection from thick disc structures, but they are probably of this kind, carrying a small fraction only of the released power. On the other hand, 2.5D MHD simulations of alpha discs recover all of the results given by self-similar solutions (see \citealt{tzef09} and references therein). 

Finally, \cite{petr10} solved the JED energy equation and showed that JEDs also harbor the three characteristic thermal solutions: two stable (a cold, geometrically thin and optically thick solution and a hot, thicker and optically thin one) and one intermediate unstable solution. It seems therefore that JEDs could account for most of the successes of the ADAFs, while explaining jet formation.

\section{JEDs in LMXBs}

LMXBs are known to exhibit hysteresis cycles which cannot be explained by the sole variation of the disc accretion rate $\dot M_a(t)$. Hysteresis is a well known phenomenon that requires the variation of at least two independent parameters. \citet{ferr06a} argue that it could be the large scale magnetic field $B_z$, or magnetic flux $\Phi$ available in the disc. Indeed, LMXBs show cycles where they switch from low luminosity hard states with a jet signature (radio emission) to high soft states with a sudden radio quenching. This can be interpreted as cycles between a JED (in its optically thin, hot branch) and a Standard Accretion Disc (hereafter SAD). Now, what determines the change from one accretion solution to another is mainly the disc magnetization $\mu \propto \Phi^2/\dot M_a$. While JEDs exist only for a narrow interval in $\mu$, the disc accretion rate is seen to vary for at least two orders of magnitude. Within our framework, this readily means that the disc magnetic flux $\Phi$ must be readjusted in the disc, implying a local variation of the field strength and/or of the JED radial extent \citep{petr08}. 

There are various processes that could lead to a variation of the vertical $B_z(r,t)$ field distribution: (1) dynamo action, (2) variation of the field polarity provided by the donor star \citep{tagg04} and (3) the interplay between turbulent diffusion and advection. The first two possibilities are of course reasonable but would be somewhat too parametric yet. The third one can be computed in the framework of alpha discs. \citet{lubo94a} showed that a SAD would get rid off its magnetic field unless the turbulent viscosity is much larger than the diffusivity (precisely, ${\cal P}_m \sim r/h$). This is kind of problematic as our current view of turbulence would suggest instead ${\cal P}_m \sim o(1)$ \cite{lesu09}.
However, they did not take into account the disc vertical equilibrium and, as pointed out by \citet{roth08}, advection could still be occurring. In fact, what is requested in order to switch from an outer SAD to an inner JED is only to reach a large enough magnetization $\mu$. Above a threshold, jets can be triggered and the field no longer diffuses away.  

Murphy, Ferreira \& Zanni (in prep) have made long term 2.5D MHD numerical simulations of an alpha disc threaded by a large scale magnetic field (for details see \citealt{murp10}). It turns out that the steady-state field distribution in the outer SAD is very steep, leading to an increase of the magnetization towards the center. However, this process takes a long time, of the order of the accretion time scale. This is potentially of large importance as the hysteresis cycles observed in LMXBs are very long and, in any case, on much longer time scales than the dynamical time scales inferred. {\em This implies that the disc magnetic field distribution $B_z(r)$ is never in a steady-state}. As a consequence, it is affected by both the initial and boundary (donor) conditions. This offers a possibility to explain why the same object can display different cycles while in outburst.

To conclude, it seems that the presence of a large scale vertical magnetic field is an unavoidable ingredient for jet launching, regardless of the nature of the central object, and a key parameter to explain LMXB long term variations.

\end{document}